\documentclass[universe,article,accept,moreauthors,10pt,a4paper]{mdpi} 

\firstpage{1} 
\articlenumber{x}
\doinum{10.3390/------}
\pubvolume{4}
\pubyear{2018}
\copyrightyear{2018}
\history{Received: 28 December 2017; Accepted: 7 February 2018; Published: date}
\pdfoutput=1

 \theoremstyle{mdpi}
 \newcounter{thm}
 \newcounter{ex}
 \newcounter{re}
\usepackage{amsfonts}
\newcommand{\pd}{\partial}
\newcommand{\goe}{\mathfrak{e}}
\newcommand{\SL}{SL(4,\mathbb R)}
\usepackage{booktabs} 
\usepackage{multirow}
\usepackage{soul} 
\usepackage{microtype}

\Title{Conformally 
Coupled General Relativity}

\Author{Andrej Arbuzov $^{1,}$* and Boris Latosh $^{2}$} 
\AuthorNames{Andrej Arbuzov and Boris Latosh}

\address{%
$^{1}$ \quad Bogoliubov Laboratory for Theoretical Physics, JINR, Dubna 141980, Russia\\
$^{2}$ \quad Dubna State University, Department of Fundamental Problems of Microworld Physics, Universitetskaya str. 19, Dubna 141982, Russia; latosh.boris@gmail.com}

\corres{Correspondence: arbuzov@theor.jinr.ru}

\abstract{The gravity model developed in the series of papers \cite{Arbuzov:2009zza,Arbuzov:2010fz,Pervushin:2011gz} is revisited. The model is based on the Ogievetsky theorem, which specifies the structure of the general coordinate transformation group. The theorem is implemented in the context of the Noether theorem with the use of the nonlinear representation technique. The canonical quantization is performed with the use of reparametrization-invariant time and Arnowitt-Deser-Misner foliation techniques. Basic quantum features of the models are discussed. Mistakes appearing in the previous papers are corrected.}


\begin{document}



\section{Introduction}\label{introduction}

General relativity forms our understanding of spacetime. It is verified by the Solar System and cosmological tests \cite{Will:2014kxa,Ade:2015xua}. The recent discovery of gravitational waves provided further evidence supporting the theory's viability in the classical regime \cite{Abbott:2016blz,Abbott:2016nmj,Abbott:2017vtc,TheLIGOScientific:2016src}. Despite these successes, there are reasons to believe that general relativity is unable to provide an adequate description of gravitational phenomena in the high energy regime and should be either modified or replaced by a new theory of gravity \cite{Nojiri:2010wj,Nojiri:2017ncd,Berti:2015itd,Clifton:2011jh,Capozziello:2011et}. 

One of the main issues is the phenomenon of inflation. It appears that an inflationary phase of expansion is necessary for a self-consistent cosmological model \cite{Linde:2007fr,Mukhanov:2005sc,Gorbunov:2011zzc}. Inflation takes place at the early stage of a universe's expansion, in the high energy regime, and should be driven either by a new scalar field \cite{Linde:1983gd} or by quantum effects associated with high order gravitational terms \cite{Starobinsky:1980te}. In that regime, gravity should be {treated} in a framework of quantum theory, but attempts to construct a {renormalizable} description within a perturbative approach have failed. It is possible to construct renormalizable gravity theory with higher derivatives \cite{Stelle:1976gc}, but general relativity is non-renormalizable in the second order of perturbation theory \cite{Goroff:1985th}. Multiple attempts to construct a proper quantum gravity model gave birth to several major research programs such as asymptotic safety \cite{Niedermaier:2006wt}, loop quantum gravity \cite{Rovelli:1997yv} and~string theory \cite{Green:1987sp,Polchinski:1998rq}.

In this paper, we revisit another approach to quantum gravity, which we are going to call Conformally-Coupled General Relativity (CCGR). CCGR is founded in a series of techniques, which may resolve some of general relativity's issues appearing in the quantum regime. These are techniques for nonlinear symmetry representation (used to implement the Ogievetsky theorem \cite{Ogievetsky:1973ik})\cite{Coleman:1969sm,Callan:1969sn,Volkov:1973vd}, the Hamiltonian formalism \cite{Arnowitt:1962hi, Dirac:1958sc}, conformal coupling of matter to gravity \cite{Dirac:1973zw, Dirac:1973gk, Deser:1970hs} and parameterization-invariant time \cite{Dirac:1950pj,Khvedelidze:1997zh,Pervushin:1999mq,Gyngazov:1998ht}. This approach was presented in a broad series of papers \cite{Arbuzov:2009zza,Arbuzov:2010fz,Pervushin:2011gz}, and our main aim is to present it in a self-consistent comprehensive form. We present a brief discussion of all aforementioned techniques in order to provide a necessary theoretical background required for CCGR. The main focus of this paper is the quantum feature of CCGR, because this was discussed very briefly and obscured in previous papers.

This paper is organized as follows. In Section \ref{motivation}, we provide a detailed motivation for the usage of these techniques, their physical relevance and applications. In Section \ref{nonlinear_symmetry_representation}, we recall the results of nonlinear symmetry realization theory. In Section \ref{nonlinear_symmetry_representation_gravity}, we demonstrate the nonlinear realization theory implementation in the context of the Ogievetsky theorem. Finally, in Section \ref{conformally_coupled_general_relativity}, we apply all the aforementioned techniques to general relativity to construct CCGR and discuss its quantum features in Section \ref{quantum_CCGR}.

\section{Motivation}\label{motivation}

The Ogievetsky theorem \cite{Ogievetsky:1973ik} is crucial for gravity research, as it specifies the structure of general coordinate transformations. Transformations are given by the following expression:
\begin{eqnarray}\label{general_coordinate_transformation}
x'_\mu = {x'}_\mu (x_0,x_1,x_2,x_3) ~,
\end{eqnarray}
where $x_\mu$ are the original coordinates, $x'_\mu$ are the new coordinates and Greek indices take their values from zero to three. One can expand \eqref{general_coordinate_transformation} in the Taylor series and obtain an infinite number of generators:
\begin{eqnarray}\label{general_coordinate_transformation_generator}
L_\mu^{n_0,n_1,n_2,n_3,} = -i x_0^{n_0} x_1^{n_1} x_2^{n_2} x_3^{n_3} \pd_\mu ~.
\end{eqnarray}

The theorem states that every generator \eqref{general_coordinate_transformation_generator} is a linear combination {of commutators formed by generators} from special linear $\SL$ and conformal $C(1,3)$ groups. Corresponding algebras are presented in Appendix \ref{appendix_algebra}. Thus, transformation features {associated with coordinate transformation} can be expressed with a finite number of generators.

A gravity model must be stated in a coordinate independent way; thus, one should be able to identify structures of $\SL$ and $C(1,3)$ groups in any gravity model.~At the same time, Noether theorems establish a connection between the model's symmetries and conservation laws, but one finds no conservation laws corresponding to the whole $\SL$ and $C(1,3)$ groups in nature. Energy--momentum and angular momentum conservation laws correspond only to the Lorentz~subgroup.

That apparent contradiction was resolved in the paper by Borisov and Ogievetsky \cite{Borisov:1974bn}. Their idea was to use the nonlinear symmetry representation theory \cite{Coleman:1969sm,Callan:1969sn,Volkov:1973vd}. Because of the nonlinear structure of the representation, one does not simply imply Noether theorems. Despite the fact that $C(1,3)$ and $\SL$ symmetries are presented, only the Lorentz subgroup receives a linear representation; the {model receives} the Lorentzian currents, but has no conservation laws for the rest of the symmetry. In the paper, the vierbein (tetrads) formalism was used to construct the representation, so such a method can be applied to any gravity model, which considers physical spacetime as a Riemannian manifold. The paper thereby provides general grounds to incorporate the Ogievetsky theorem within any gravity model in a four-dimensional (Riemannian) spacetime.

The benefits of this method appear at the level of quantum theory,
as it introduces new symmetry, which may influence the model's behavior. New symmetry may exclude certain terms from a Lagrangian, thereby improving the renormalization behavior.

Another important issue is the definition of a gravity coupling to matter. The coupling should be driven by some fundamental physical principle, as this allows one to narrow down a number of possible models.
An example of a similar approach is given by the Lovelock theorem \cite{Lovelock:1971yv}. It proves that the left-hand side of Einstein equation
is fixed uniquely, if one considers it to be a second differential order second-rank 
 divergenceless tensor. Another example is Horndeski's paper \cite{Horndeski:1974wa}, where it was proven that scalar-tensor models with a nontrivial kinetic coupling can have second differential order field equations, {but there are only two suitable interaction terms.}
In such a way, it appears to be fruitful to find a guiding physical principle constraining a number of possible interactions between matter and~gravity.

A principle {that may help one define the gravitational coupling in a unique way} was proposed in the papers of Dirac and Deser \cite{Dirac:1973zw,Dirac:1973gk,Deser:1970hs}. The paper by Dirac follows the large numbers hypothesis \cite{Dirac:1938mt}, which states that large dimensionless physical constants are connected to a cosmological epoch. Dirac proposed considering matter coupled to a metric $\tilde g$, which differs from the full metric $g$ governed by Einstein equations \cite{Dirac:1973gk}. Therefore, the full metric $g$ cannot be measured directly in any experiment, but~it defines particles' motion; the metric $\tilde g$ is measured by our apparatus, but does not define geodesic motion. The paper by Deser \cite{Deser:1970hs} is devoted to the study of massless scalar fields, which are, unlike massless vector and spinor fields, not coupled to gravity in a conformally-invariant way. He proved that one can introduce additional terms to the standard massless scalar field Lagrangian, which restore the invariance. The resulting action matches the standard Hilbert action up to a conformal transformation. Thus, such a construction separates a conformal gravitational scalar degree of freedom from the full~metric.

These results can be used in the following way. First, following Dirac's idea, one constructs a coupling of matter to gravity in a conformally-invariant way. Second, following Deser's paper, one separates the conformal metric at the level of an action via conformal transformations. Thus, an action turns out to be separated into two parts: one drives a cosmological evolution and the other drives local gravitational interaction. To the best of our knowledge, Penrose, Chernikov and Tagirov \cite{penrose_1964,Chernikov:1968zm} were the first to write the Hilbert action {in such a form}. Usage of the conformal gravity coupling gave birth to conformal cosmology \cite{Behnke:2001nw,Blaschke:2004by,Barbashov:2005ne,Zakharov:2010nf}.

We would like to highlight that the conformal symmetry plays an important role in physics. The best example illustrating its role is the standard model of particle physics, as it appears to be almost conformally invariant. Without the Higgs sector, the model admits the conformal invariance, because it consists of massless vector bosons and fermions. Higgs boson is coupled to other particles in a conformally-invariant way with the use of dimensionless couplings. The conformal invariance is violated when one introduces the scale of spontaneous symmetry breaking. Moreover, as one introduces a single scale, it cannot be evaluated from the first principles, and one must refer to experimental data to establish it.

We imply the Hamiltonian formalism, as it is considered to be the canonical approach to quantum theory. The formalism was developed by Arnowitt, Deser and Misner \cite{Arnowitt:1962hi} (and founded by Dirac~\cite{Dirac:1958sc}). It treats a four-dimensional spacetime as a series of three-surfaces with dynamical geometry. 
The main implication of the Hamiltonian formalism in this paper is reduced to usage of the ADM (Arnowitt-Deser-Misner) foliation technique for the sake of simplicity. 

The implication of ADM 
foliation gives rise to the time parameterization problem that was addressed in the papers \cite{Khvedelidze:1997zh,Pervushin:1999mq,Gyngazov:1998ht}. Within quantum theory, time is treated as an external parameter mapping the system's evolution. In the realm of gravity models, time is merely a coordinate that can be chosen arbitrarily. In the realm of cosmology, one expects to associate time with a cosmological epoch, which can be defined via cosmological parameter measurement. Therefore, one expects to connect the time with an observable quantity and should find a way to reduce an ambiguity from the definition of time. In the series of papers~\cite{Khvedelidze:1997zh,Pervushin:1999mq,Gyngazov:1998ht}, a method was proposed that allows one to define time in a reparametrization-independent way and to relate it to a cosmological epoch according to the Einstein cosmological principle \cite{Einstein:1917ce}.

The motivation should be summarized as follows. First, we use nonlinear representation of $\SL$ and $C(1,4)$ groups to incorporate the Ogievetsky theorem into the model. Second, we~introduce a conformal coupling of matter to gravity and separate the conformal metric at the level of an action. Next, we implement ADM foliation to define time in a reparametrization invariant way. Techniques used to obtain the classical model affect the quantum version of CCGR, namely additional symmetry allows one to introduce the notion of (conformal) gravitons, for which the Lagrangian is bilinear. Nonetheless gravity preserves its nonlinear nature, as gravitational interaction cannot be reduced to conformal gravitons only. However, the implication of all aforementioned techniques may improve the study of CCGR quantum properties.

\section{Nonlinear Symmetry Representation}\label{nonlinear_symmetry_representation}

Nonlinear symmetry representation theory is {presented} in the papers \cite{Coleman:1969sm,Callan:1969sn,Volkov:1973vd}. One can refer to the papers~\cite{Ivanov:2016lha,Goon:2014paa,Goon:2012dy} for a detailed review; a detailed example of a model with a nonlinear representation of $SO(N)$ on the $SO(N)/SO(N-1)$ coset in the context of the nonlinear sigma model is given in the textbook~\cite{Nair:2005iw}.

One starts with a $d$-dimensional Lie group $G$ and its $n$-dimensional subgroup $H$. Lie algebra of $H$ is formed by generators $V_l$, $(l=1,\cdots,n)$; Lie algebra of $G$ is formed by generators from algebra H
 and $A_l$, $(l=1,\cdots,d-n)$. An arbitrary element $g$ from a neighborhood of the identity of $G$ can be presented in the following form:
\begin{eqnarray}\label{arbitrary_group_element_parameterization}
g(\zeta,\varphi) = e^{\zeta_l A_l} ~ e^{\varphi_l V_l} ~.
\end{eqnarray}

Element $\exp[\varphi_l V_l]$ belongs to a neighborhood of the identity of $H$; element $\exp[\zeta_l A_l]$ belongs to the $G/H$ coset. Parameterization \eqref{arbitrary_group_element_parameterization} sets coordinates on group $G$, group $H$ and the $G/H$ coset.

The coset is formed by right equivalence classes of $G$ with respect to $H$. Elements $g$ and $g'$ lie in the same right equivalence class of $G/H$, if and only if $h \in H$ exists such that $g' =g h$. In~parameterization~\eqref{arbitrary_group_element_parameterization}, all equivalent (with respect to H
) elements of $G$ have the same coset coordinates~$\zeta_l$.

Group $G$ has a natural left group action on itself. An element $g\in G$ acts on an element $g' \in G$ by a right multiplication as $g g'$. One can narrow this group action down to the $G/H$ coset. The action of an arbitrary element $g$ from $G$ on an arbitrary element $\exp[\zeta_l A_l]$ from $G/H$ is given by the following:
\begin{eqnarray}\label{coset_transformation_formula}
g ~e^{\zeta_l A_l} = e^{\zeta'_l A_l} e^{u_l V_l} ~,
\end{eqnarray}
where $\zeta_l$ are coordinates of the initial coset element, $\zeta'_l(\zeta,g)$ are coordinates of the element after the group action and $u_l(\zeta,g)$ are coordinates on $H$ that form an associated nonlinear representation of $G$ on $H$.

In such a way, $\zeta'_l(\zeta,g)$ is a non-linear representation of $G$ on $G/H$, and $u_l(\zeta,g)$ is a nonlinear representation of $G$ on $H$. In Formula \eqref{coset_transformation_formula}, if $g$ lies in $H$, then the representation $\zeta'_l (\zeta,g)$ acts as an adjoint representation of $H$ on $G$, and $u_l(\zeta,g)$ becomes a linear representation of $H$. Therefore, one can consider nonlinear representations generated by \eqref{coset_transformation_formula} as a spontaneous breaking of the symmetry group $G$ down to $H$.

To relate these representations with physical fields, one should, first, treat coset coordinates $\zeta_l$ as physical fields subjected to the group transformation \eqref{coset_transformation_formula}: $\zeta_l \to \zeta'_l (\zeta,g)$. Second, one should take physical fields $\psi_i$ subjected to a linear representation of H: $\psi \to \hat{D}\left( e^{\phi_l V_l} \right) \psi$; and use $u_l(\zeta,g)$ defined by \eqref{coset_transformation_formula}, except transformation parameters $\phi_l$. Finally, one should introduce a covariant derivative $\nabla$ that transforms in agreement with the nonlinear representation. Covariant derivatives are given by the~following:
\begin{eqnarray}
 \nabla_\mu \psi &=& \pd_\mu \psi + v_\mu ~\hat T \psi ~, \\
 \nabla_\mu \zeta &=& p_\mu ~,
\end{eqnarray}
where values of $p_\mu$ and $v_\mu$ are defined by the form:
\begin{eqnarray}
e^{-\zeta_l A_l} \pd_\mu e^{\zeta_l A_l} = p_\mu A_\mu + v_\mu V_\mu ~.
\end{eqnarray}

One should use fields $\zeta$, $\psi$ and their covariant derivatives to construct an invariant Lagrangian. 

A nonlinear symmetry representation, as we mentioned before, can be treated as spontaneous symmetry breaking, so a correspondent model is constrained by the Goldstone theorem. Therefore, $d-n$ physical fields associated with coset coordinates can only be massless Goldstone particles. As a nonlinear representation of $G$ on $H$ is founded on a linear representation of $H$, one can introduce an arbitrary number of additional (massive or massless) degrees of freedom subjected to various linear representations of $H$.

The nonlinear representation implementation appears to be fruitful in the quantum regime. 
It is impossible to distinguish a model with a linear representation of $H$ and non-minimal interaction from a model with a nonlinear representation of $G$ and minimal interaction solely by the form of the classical Lagrangian. In the quantum regime, symmetry may prevent some interactions from appearance, which is crucial for renormalization features of the model and for its physical content. This is the main reason why we are interested in the usage of that technique, and we discuss this issue in detail in Section \ref{quantum_CCGR}.

\section{Nonlinear Representation of $\SL$ and $C(1,3)$ on the Vierbein}\label{nonlinear_symmetry_representation_gravity}

The technique used to describe gravity with a nonlinear representation of $\SL$ and $C(1,3)$ was proposed in the paper \cite{Borisov:1974bn}. A more detailed review can be found in \cite{Ivanov:2016lha}. In accordance with Section \ref{nonlinear_symmetry_representation}, one should use variables subject to a linear Lorentz transformation to construct nonlinear representations of $\SL$ and $C(1,3)$. Vierbeins are objects that one should use for that purpose~\cite{Chandrasekhar:1985kt,Misner:1974qy,Kiefer:2004gr}. We recall the vierbein formalism and then turn to a nonlinear $\SL$ and $C(1,3)$ group representation.

There are several ways to construct the vierbein formalism. One can introduce the vierbein to obtain a proper generalization of Dirac's equation. In the realm of flat spacetime, all physical quantities
transform by Lorentz group representations. In the realm of curved spacetime, an arbitrary tensor transforms in accordance with the general linear transformation group $\SL$. However, there is no transformation in $\SL$ that corresponds to the spinor Lorentz group representation. One~introduces vierbein $\goe_{\mu a}$, which transforms in accordance with $\SL$ by the Greek index and in accordance with the Lorentz group by the Latin index (both Latin and Greek indices take values from zero to three in this section). These objects connect the spinor Lorentz group representation with the $\SL$ representation. To generalize Dirac's equation one simply replaces the index of standard $\gamma$-matrices:
\begin{eqnarray}
\gamma_\mu = \goe_{\mu}^{~~m} \gamma_m.
\end{eqnarray}

In the same manner, one uses the vierbein to replace $\SL$ indices with Lorentz ones in an arbitrary~tensor.

An alternative approach to the vierbein considers vector fields on a spacetime. The set of all vector field forms a linear space, and one is free to choose its basis. The standard choice is the holonomic basis; these are vector fields co-directed with coordinate lines in every point of a spacetime. This basis is canonically denoted as $\pd_\mu$( and $dx^\mu$), because these derivatives transform as vectors \cite{Chandrasekhar:1985kt,kuhnel2015differential}. The manifold metric can be treated as a bilinear form on the linear space of vector fields:
\begin{eqnarray}
g_{\mu\nu} ~ dx^\mu dx^\nu \leftrightarrow g_{\mu\nu} ~dx^\mu \otimes dx^\nu = g^{\mu\nu} ~ \pd_\mu \otimes \pd_\nu ~.
\end{eqnarray}

One can choose a vector field basis $\omega_m$, which makes the metric take the Minkowski form:
\begin{eqnarray}
g^{\mu\nu} ~\pd_\mu \otimes \pd_\nu= \eta^{mn} ~\omega_m \otimes \omega_n ~,
\end{eqnarray}
where $\eta_{mn}$ is the Minkowski metric and $\omega_m$ is the vierbein basis (of a vector field space). We should highlight that the choice of a basis is not related to a coordinate frame on a spacetime. It corresponds to gauge transformations or to a coordinate transformation in the tangent bundle. Coefficients that relate the holonomic basis and forms $\omega_m$ are called vierbeins:
\begin{eqnarray}
\pd_\mu = \goe_{\mu}^{~~m} \omega_m ~.
\end{eqnarray}

In the vierbein basis, the metric tensor coincides with the Minkowski one, and as the form of the metric should be preserved, vector fields $\omega_m$ can only be subjected to Lorentz transformations. Thus, the vierbein transforms by the Lorentz group by the Latin index and in accordance with $\SL$ by the Greek index. The differential geometry formalism can be stated in terms of the vierbein \cite{Chandrasekhar:1985kt,kuhnel2015differential}, and in Appendix \ref{appendix_vierbein} we present a set of formulae that relate the standard and vierbein formalisms.

Now, we turn to the discussion of nonlinear representations of $\SL$ and $C(1,3)$. For the sake of simplicity, we consider the affine group $A(4)$ instead of $\SL$. This choice has no influence on the representation, because we did not include either new generators or new structures. The affine group consists of all linear transformations in a four-dimensional spacetime $x'_\mu= a_\mu^{~\nu} x_\nu+c_\mu$; it is a semidirect product of $\SL$ and a group of all shifts. Following the paper \cite{Borisov:1974bn}, first, we consider a nonlinear representation of $A(4)$ on the $A(4)/L$ coset. Group $A(4)$ consists of shift generators $P_{(\mu)}$, Lorentz group (asymmetric) generators $L_{(\mu)(\nu)}$ and distortion (symmetric) generators $R_{(\mu)(\nu)}$. We~use indices in brackets to numerate generators of the algebra, as they are not (yet) related to any representation of the Lorentz group. Greek indices with and without brackets take their values from zero to three. The correspondent commutation relations are given in Appendix \ref{appendix_algebra}.

An arbitrary element $G$ from the $A(4)/L$ coset is parameterized as follows:
\begin{eqnarray}\label{the_main_coset_parameterization}
G(x,h)= \exp\left[i x_{(\mu)} P_{(\mu)}\right] ~\exp\left[ \cfrac{i}{2} ~h_{(\mu)(\nu)} R_{(\mu)(\nu)}\right] ~.
\end{eqnarray}

The following forms define transformation features of all objects in the representation:
\begin{eqnarray}
G^{-1} d G &=&i \left( \omega_{(\mu)} P_{(\mu)} + \omega^L_{(\mu)(\nu)} (d) ~L_{(\mu)(\nu)} + \omega^R_{(\mu)(\nu)} (d) R_{(\mu)(\nu)}\right) ~, \\
 \omega_{(\mu)}&=& \omega_{(\mu)\nu} dx^\nu ~, \label{form_omega} \\
 \omega^R_{(\mu)(\nu)} (d) &=&1/2~ \left( \omega_{(\mu)}^{~~~\sigma} d\omega_{(\nu)\sigma} + \omega_{(\nu)}^{~~~\sigma} d\omega_{(\mu)\sigma} \right) ~, \label{omega^R} \\
 \omega^L_{(\mu)(\nu)} (d)&=&1/2~ \left( \omega_{(\mu)}^{~~~\sigma} d\omega_{(\nu)\sigma} - \omega_{(\nu)}^{~~~\sigma} d\omega_{(\mu)\sigma} \right) ~. \label{omega^L}
\end{eqnarray}

As we wrote before, $L_{(\mu)(\nu)}$ is asymmetric, and $R_{(\mu)(\nu)}$ is symmetric; therefore, the form $\omega^L$ is also asymmetric, and the form $\omega^R$ is symmetric. One should treat forms $\omega_{(\mu)}$ as a vector field basis on a spacetime and $\omega_{(\mu)\nu}$ as the vierbein:
\begin{eqnarray}
g_{\mu\nu} dx^\mu \otimes dx^\nu = \eta_{(\mu)(\nu)} \{ \omega_{(\mu)\alpha} dx^\alpha \} \otimes \{ \omega_{(\nu)\beta} dx^\beta \}= \eta_{(\mu)(\nu)} ~\omega_{(\mu)} \otimes \omega_{(\nu)} ~.
\end{eqnarray}

The same thing should be done with a conformal group representation on the $C(1,3)/L$ coset. For the sake of simplicity, we do not present the procedure here, as it can be found in the original paper \cite{Borisov:1974bn}. These nonlinear representations should agree with one another to provide the following definition of a covariant derivative (applied to a field of an arbitrary spin):
\begin{eqnarray}
 \nabla_{(\mu)} \Psi &=& \omega_{(\mu)}^{~~\nu} \pd_\nu \Psi + \cfrac{i}{2} ~V_{(\mu),(\alpha)(\beta)} ~L^\Psi_{(\alpha)(\beta)} \Psi ~,
 \label{nonlinear_representation_gravity_covariant_derivate}\\
V_{(\mu),(\alpha)(\beta)} &=& \omega^L_{(\alpha)(\beta)}(\pd_{(\mu)}) + \omega^R_{(\alpha)(\mu)}(\pd_{(\beta)}) -\omega^R_{(\beta)(\mu)}(\pd_{(\alpha)}) ~.\label{nonlinear_representation_gravity_connection}
\end{eqnarray}

Here, $L^\Psi_{(\alpha)(\beta)}$ is the Lorentz group representation operator acting on $\Psi$, and forms $\omega^L$ and $\omega^R$ are given by \eqref{omega^R} and \eqref{omega^L}; \eqref{nonlinear_representation_gravity_connection} is the spin connection. The Riemann tensor $R_{(\mu)(\nu)(\alpha)(\beta)}$ is defined in the following~way:
\begin{eqnarray}
[\nabla_{(\mu)}, \nabla_{(\nu)}] \psi_A = \cfrac{i}{2} R_{(\mu)(\nu)(\alpha)(\beta)} {L^\Psi}_{(\alpha)(\beta)} \Psi 
\end{eqnarray}
and given by the following formula:
\begin{eqnarray}\label{Riemann_tensor_formula}
R_{(\mu)(\nu)(\alpha)(\beta)} = \pd_{(\mu)} V_{(\nu),(\alpha)(\beta)} + V_{(\mu),(\nu)(\tau)} V_{(\tau),(\alpha)(\beta)} - \left[ (\mu)\leftrightarrow (\nu) \right] ~.
\end{eqnarray}

The Riemann tensor \eqref{Riemann_tensor_formula} is bilinear in $\omega^R$, $\omega^L$ and linear in their derivatives, because of the form of the spin connection \eqref{nonlinear_representation_gravity_connection}.

Vierbein $\omega_{(\mu)\nu}$ and forms $\omega^R$ thereby should be treated as dynamical variables of the theory, as~they carry system symmetry. Despite the fact that both forms $\omega^R$ and $\omega^L$ enter a covariant derivative \eqref{nonlinear_representation_gravity_covariant_derivate}, only forms $\omega^R$ can change during the system's evolution. Definitions \eqref{omega^R} and \eqref{omega^L} result in the following identity:
\begin{eqnarray}
 \omega_{(\mu)}^{~~~\sigma} d\omega_{(\nu)\sigma} = \omega^R_{(\mu)(\nu)} (d) + \omega^L_{(\mu)(\nu)} (d) ~.
\end{eqnarray}

As vierbeins are constrained by orthogonality identities (see \eqref{orthogonality_vierbein} in Appendix \ref{appendix_vierbein}), the following expression holds for a vierbein differential:
\begin{eqnarray}
 d\omega_{(\nu)\rho}= \omega_{(\mu)\rho} \left( \omega^R_{(\mu)(\nu)} + \omega^L_{(\mu)(\nu)} \right).
\end{eqnarray}

This identity should be used to obtain the following expression for a metric differential:
\begin{eqnarray}
 d g_{\mu\nu} = d(\omega_{(\sigma)\mu} \omega_{(\sigma)\nu})= \left( \omega_{(\rho)\mu} \omega_{(\sigma)\nu} + \omega_{(\sigma)\nu} \omega_{(\rho)\mu} \right) \left(\omega^R_{(\rho)(\sigma)} (d) + \omega^L_{(\rho)(\sigma)} (d)\right).
\end{eqnarray}

The forepart $\left( \omega_{(\rho)\mu} \omega_{(\sigma)\nu} + \omega_{(\sigma)\nu} \omega_{(\rho)\mu} \right)$ is symmetric with respect to indices $(\mu)$ and $(\nu)$, while the former part contains an asymmetric form $\omega^L_{(\mu)(\nu)}$. {Hence, the asymmetric form $\omega^L$ does not contribute to the differential because of the symmetry features}:
\begin{eqnarray}
 dg_{\mu\nu} = \left( \omega_{(\rho)\mu} \omega_{(\sigma)\nu} + \omega_{(\sigma)\nu} \omega_{(\rho)\mu} \right) \omega^R_{(\sigma)(\rho)}(d) ~.\label{metric_differential_forms}
\end{eqnarray}

As the form $\omega^L$ does not enter the metric differential, it does not evolve and cannot be considered as a dynamical variable. 

We would like to highlight the following feature of the representation. One can construct a nonlinear representation of $A(4)$ on the $A(4)/L$ coset, but in that case, a covariant derivative cannot be determined in a unique way \cite{Borisov:1974bn}. 
One excludes such an ambiguity by combing representations of $A(4)$ and $C(1,3)$. On the classical level, the technique 
does not introduce new additional degrees of freedom, and at the level of the classical Lagrangian, the model is indistinguishable from general relativity. At~the same time, the construction presented in this section does not agree with the conformal coupling condition; the technique merely connects a nonlinear representation to a vierbein. We introduce a conformally-invariant coupling in the next section.

\section{Conformally-Coupled General Relativity}\label{conformally_coupled_general_relativity}

Following the motivation, presented in Section \ref{motivation}, we apply the discussed techniques to general relativity. {They allows us to achieve two goals: to introduce new symmetry in the model and to choose a time variable that may be associated with an observable quantity. In this section, we show that one can indeed implement the aforementioned techniques and formulate a gravity action in terms of conformal variables and parameterization-invariant time. The action obtained at the end of the section allows one to study new features of CCGR that do not appear in general relativity. In the next section, we use it to construct a correspondent quasiclassical quantum model.}

First, we separate the conformal degree of freedom from the full metric:
\begin{eqnarray}\label{metric_decomposition}
g_{\mu\nu} ~dx^\mu \otimes dx^\nu = e^{-2 D} ~\widetilde{g}_{\mu\nu} ~d\chi^\mu \otimes d\chi^\nu.
\end{eqnarray}

Here, $g_{\mu\nu}$ in the full (Einstein) metric, $\widetilde{g}_{\mu\nu}$ in the conformal metric, $x_\mu$ are arbitrary coordinates on a spacetime, $\chi_\mu$ are correspondent conformally-invariant coordinates and $D$ is the conformal degree of freedom, also known as the dilaton. Metric parameterization \eqref{metric_decomposition} results in the Penrose--Chernikov--Tagirov (PCT) action \cite{penrose_1964,Chernikov:1968zm}. {At this level, we impose the conformal coupling to matter and obtain the following action}:
\begin{eqnarray}
& S_\text{Hilbert} = \int d^4 x \sqrt{-g} \left[ \cfrac{M_P^2}{16\pi} (R-2\Lambda) + L_\text{matter} (g_{\mu\nu}) \right] \to \label{Hilbert_action} \\
& S_\text{CCGR}=\int d^4 \chi \sqrt{-\widetilde g} \left[\cfrac{\widetilde{M}_P^2}{16\pi} \left( \widetilde R - 2 \widetilde \Lambda \right) + \cfrac{3 \widetilde{M}_P^2}{8\pi} \left( \widetilde{g}^{\mu\nu} \nabla_\mu D \nabla_\nu D \right) + L_\text{matter}(\widetilde{g}_{\mu\nu}) \right] .\label{conformal_action}
\end{eqnarray}

Here, $\Lambda$ is the cosmological constant, $\widetilde{\Lambda} = e^{-2D} \Lambda$ is the conformal cosmological constant, $M_P$ is the Planck mass and $\widetilde{M}_P=M_P e^{-D}$ is the conformal Planck mass. {The action \eqref{conformal_action} is not equivalent to the Hilbert action \eqref{Hilbert_action} and does not describe general relativity. The reason is in the conformal gravitational coupling. The Hilbert action \eqref{Hilbert_action} describes matter coupled to the full (Einstein) metric, which does not admit conformal invariance, while the CCGR action \eqref{conformal_action} preserves the coupling invariant under conformal transformations. In such a way, at this level, we do not consider general relativity, but its modification}. We introduce the cosmological constant in the model for the sake of generality and treat it as a free model parameter. However, verification by Type Ia supernovae data showed that a conformal cosmological model based on the action \eqref{conformal_action} {with a vanishing cosmological constant provides a good fit for cosmological data and} can be treated alongside the standard cosmological model~\cite{Zakharov:2010nf}.

We define the (conformal) vierbein by the following:
\begin{eqnarray}
 g_{\mu\nu} dx^\mu \otimes dx^\nu= e^{-2 D} ~\widetilde{g}_{\mu\nu} ~d\chi^\mu \otimes d\chi^\nu = e^{-2D} \eta_{(\mu)(\nu)} ~ \omega_{(\mu)} \otimes \omega_{(\nu)}.
\end{eqnarray}

Here, $\omega_{(\mu)}$ is the (conformal) vierbein basis subjected to the nonlinear symmetry representation \eqref{form_omega}. {Because of the new symmetry, one can operate with a series of equivalent vierbeins related by the symmetry transformation.} The choice of vierbein corresponds to different frames in \eqref{the_main_coset_parameterization} and does not affect the structure of \eqref{the_main_coset_parameterization} or the structure of the representation.

The conformal metric is conformally invariant, so any conformal transformation $\widetilde{g}_{\mu\nu} \to e^{2\Omega} \widetilde{g}_{\mu\nu}$ are equipped with a dilaton transformation $D \to D + \Omega$ preserving the form of the metric \eqref{metric_decomposition}. In terms of the (conformal) vierbein, conformal symmetry is given by the following:
\begin{eqnarray}\label{conformal_symmetry_on_vierbein}
 \begin{cases}
 \omega_{(\mu)} \to e^{\Omega} \omega_{(\mu)} ~,\\
 D \to D + \Omega .
 \end{cases}
\end{eqnarray}

The conformal metric and the dilaton are dependent variables, because {we did not introduce new degrees of freedom in transition from \eqref{Hilbert_action} to \eqref{conformal_action}.} We relate the dilaton to the three-metric ${}^{(3)} g_{mn}$ defined by the full metric $g_{\mu\nu}$ as follows:
\begin{eqnarray}
 D = -\cfrac16 \ln\det {}^{(3)} g_{mn} ~. \label{dilaton_and_three_metric}
\end{eqnarray}

One cannot express the dilaton through the conformal metric because of the symmetry \eqref{conformal_symmetry_on_vierbein}. Instead, one can fix a gauge of variables $\widetilde g$ and $D$, in full analogy with the standard gauge field theory.

Thus, we implemented a nonlinear symmetry realization and conformal coupling {to general relativity and constructed CCGR. Formally, the gravitational sector of CCGR is indistinguishable from general relativity at the level of the classical action. The difference appears because of the new symmetry of dynamical variables and conformal coupling to matter. In the rest of the section, we introduce a parametrization-invariant time. This allows us to separate global and local dynamic of gravitational field, which may be viewed as one of the main results of CCGR.} 

 {At the quantum level, CCGR also differs from general relativity. While the metric components are treated as dynamic variables in general relativity, one must use the vierbein and forms $\omega^R$ in CCGR, because they carry the model's symmetry. They introduce a new way to map both system's dynamic and physical states. We discuss these issues in the following section.} 

 {In order to obtain a notion of parametrization-invariant time, one needs} to implement the Hamiltonian formalism and the ADM foliation \cite{Arnowitt:1962hi}. We split the conformal four-dimensional spacetime into a series of three-spaces equipped with a metric $\gamma_{ab}$. The conformal metric $\widetilde{g}_{\mu\nu}$ is parameterized by the three-metric, the shift three-vector $N_a$ and by the laps function $N$ (see Appendix \ref{appendix_ADM_foliation} for notations and definitions). {The laps function contains data on time parametrization, so we show a way to relate it to the dilaton. As the dilaton itself is related to the cosmological scale factor (i.e., three-metric determinant), we establish a way to relate the coordinate time to the cosmological time in a parametrization-independent way}.


In the papers \cite{Khvedelidze:1997zh,Pervushin:1999mq,Gyngazov:1998ht}, it was proposed to associate physical time with the zero dilaton mode according to the Einstein cosmological principle \cite{Einstein:1917ce}. 
We use the following definition of a three-space mean value of a physical quantity $X$, {which depends only on the time variable}:
\begin{eqnarray}
 \langle X \rangle (\chi^0) = \cfrac{1}{V_0} \int_{V_0} d^3 \chi ~\sqrt{\gamma} ~X(\chi^0,\chi^1,\chi^2,\chi^3) ~,
\end{eqnarray}
where $V$ is the three-volume of a region over which the averaging is carried out. In such a way, any physical quantity $X$ is separated into two parts: the mean part $\langle X\rangle$ and the fluctuation part $\bar X$. They are connected by the following expression:
\begin{eqnarray}
 X (\chi^0,\chi^1,\chi^2,\chi^3) = \langle X \rangle (\chi^0) + \bar X (\chi^0,\chi^1,\chi^2,\chi^3) 
\end{eqnarray}
and the following orthogonality condition holds:
\begin{eqnarray}
 \langle \bar X \rangle = 0. \label{mean_value_orthogonality_condition}
\end{eqnarray}
 
 {Thus, one can define zero dilaton mode $\langle D \rangle$, which corresponds to its global structure, and fluctuation mode $\bar D$, which corresponds to local dynamics:}
\begin{eqnarray}
D(\chi^0,\chi^1,\chi^2,\chi^3) = \langle D \rangle (\chi^0) + \bar D (\chi^0,\chi^1,\chi^2,\chi^3) ~.
\end{eqnarray}

This allows one to exclude all interferences between $\bar D$ and $\langle D \rangle$ from the action \eqref{conformal_action} and to reduce it to the following form:
\begin{eqnarray}
S&=&\int d^4 \chi \sqrt{\gamma} N \left[\cfrac{\widetilde{M}_P^2}{16\pi} \left( \widetilde R - 2 \widetilde \Lambda \right) - \cfrac{3 \widetilde{M}_P^2}{8\pi} \cfrac{1}{N^2} \left( \cfrac{\pd}{\pd\chi^0} \langle D \rangle \right)^2 \nonumber \right. \\
& & \hspace{4cm} \left. + \cfrac{3 \widetilde{M}_P^2}{8\pi}~ \left( \widetilde{g}^{\mu\nu} \nabla_\mu \bar D \nabla_\nu \bar D \right) + L_\text{matter}(\widetilde{g}_{\mu\nu}) \right] . \label{action_pre_decomposed} 
\end{eqnarray}

The zero dilaton mode action reads:
\begin{eqnarray}
S_{\langle D\rangle} = -\cfrac{3\widetilde{M}_P^2}{8\pi} \int d \chi^0 ~\left( \cfrac{\pd}{\pd \chi^0} \langle D \rangle \right)^2 ~\int_{V_0} d^3\chi \sqrt{\gamma} ~\cfrac{1}{N} ~. \label{pure_dilaton_action_1}
\end{eqnarray}

It gives a way to introduce the parameterization-invariant time. {In full analogy with the dilaton, one separate local and global parts of the lapse function N
}:
\begin{eqnarray}
 N (\chi^0,\chi^1,\chi^2,\chi^3) = N_0 (\chi^0) \mathcal N (\chi^0,\chi^1,\chi^2,\chi^3) ~,
\end{eqnarray}
where $N_0$ is defined by the following:
\begin{eqnarray}
\cfrac{1}{N_0} = \left\langle \cfrac{1}{N} \right\rangle ~ \label{N_0_definition}
\end{eqnarray}
and $\mathcal N$ is subjected to the orthogonality condition: $\langle \mathcal{N}^{-1} \rangle =1$. The dilaton action \eqref{pure_dilaton_action_1} {turns out to be defined in terms of global components only}:
\begin{eqnarray}
S_{\langle D\rangle} = - V_0 \cfrac{3\widetilde{M}_P^2}{8\pi} \int d \chi^0 N_0 ~\left( \cfrac{1}{N_0} \cfrac{\pd}{\pd \chi^0} \langle D \rangle \right)^2 \label{pure_dilaton_action} .
\end{eqnarray}

Moreover, in the action \eqref{pure_dilaton_action}, any time reparametrization $\chi^0 \to \chi'^0$ can be absorbed into the definition of $N_0$. In such a way, one can define parameterization-invariant time $t$ by the following~expression:
\begin{eqnarray}
dt = N_0 d \chi^0 ~. \label{parametrization_invariant_time}
\end{eqnarray}

Additionally, the four-volume element can be expressed as follows:
\begin{eqnarray}
d^4 x ~\sqrt{-\widetilde g} = d\chi^0 N_0 ~d^3\chi \sqrt{\gamma} ~ \mathcal N = dt ~d^3\chi \sqrt{\gamma} ~ \mathcal N .
\end{eqnarray}

  {In order to associate time with a cosmological epoch, one needs to relate parametrization-invariant variable $t$ with the zero dilaton mode $\langle D \rangle$. Action \eqref{pure_dilaton_action} provides a way to establish such a relation, because it independently introduces a correspondent} Jacobian matrix $d \langle D \rangle / N_0 d\chi^0$. Therefore, zero dilaton mode can be considered as a time variable alongside the parameterization-invariant time~\eqref{parametrization_invariant_time}.

That mechanism has two direct corollaries. First, the canonical momentum of the zero dilaton mode $P_{\langle D\rangle}$ {calculated} with respect to time variable $\langle D \rangle$ is related to the evolution operator and must be non-vanishing:
\begin{eqnarray}
P_{\langle D\rangle} = \cfrac{3 \widetilde{M}_P^2}{4\pi} ~ \cfrac{1}{N_0} \cfrac{\pd}{\pd \chi^0} ~\langle D \rangle \not =0 \label{Mean_D_momentum} ~.
\end{eqnarray}
 
 {Moreover, momentum \eqref{Mean_D_momentum} is proportional to the Jacobi matrix relating parametrization-invariant time $t$ and $\langle D \rangle$. Thus, if $P_{\langle D \rangle}$ vanished, one cannot relate parametrization-invariant time with the zero dilaton value, i.e., to the scale factor \eqref{dilaton_and_three_metric} describing the expansion of three-geometry}. Second, because of the orthogonality condition \eqref{mean_value_orthogonality_condition}, $\bar D$ does not depend on the $\langle D \rangle$, and the correspondent canonical momentum is zero:
\begin{eqnarray}
P_{\bar D} = \cfrac{1}{N} \left( \cfrac{\pd}{\pd \chi^0} - N^a \pd_a \right)~\bar D =0~.
\end{eqnarray}

 {Therefore, the CCGR action \eqref{action_pre_decomposed} consists of three parts}:
\begin{eqnarray}
S= S_\text{Universe} + S_\text{Gravitons} +S_\text{Potential} ~, \label{decoupled_Lagrangian_sum}
\end{eqnarray}
\begin{eqnarray}
S_\text{Universe} &=& -V_0 \int d \chi^0 N_0 ~\cfrac{3\widetilde{M}_P^2}{8\pi} \left( \cfrac{1}{N_0} \cfrac{\pd}{\pd \chi^0} \langle D \rangle \right)^2 ~,\label{decoupled_Lagrangian_Cosmology}\\
S_\text{Gravitons} &=& \int d\chi^0 N_0 ~ \int d^3\chi \sqrt{\gamma} ~ \mathcal N ~\cfrac{\widetilde{M}_P^2}{16\pi} \left( \widetilde R - 2 \widetilde \Lambda \right) ~, \label{decoupled_Lagrangian_Graviton}\\
S_\text{Potential} &= & \int d\chi^0 N_0 ~\int d^3\chi \sqrt{\gamma} ~ \mathcal N ~ \cfrac{3 \widetilde{M}_P^2}{8\pi}~ \gamma^{ab} \pd_a \bar D \pd_b \bar D ~. \label{decoupled_Lagrangian_Potential}
\end{eqnarray}

  {That decomposition allows us to explain new features of CCGR. The decomposition was obtained due to the implementation of the new symmetry and introduction of a conformally-invariant time coordinate $\langle D\rangle$.} It shows that the cosmological evolution is driven by the mean (global) structure of a gravitational field described by the $\langle D \rangle$; while local gravitational interaction is driven by the fluctuation part of a gravitational field $\bar D$. {In other words, we established independent parameterization of global (cosmological) and local dynamics of the gravitation field. Action \eqref{decoupled_Lagrangian_sum} is an explicit representation of this result.} Local and global degrees of freedom interact with each other in a non-trivial way, and the interaction is given by \eqref{decoupled_Lagrangian_Potential}. Corollaries of that model that take place in the classical regime were studied in the papers \cite{Arbuzov:2009zza,Arbuzov:2010fz,Pervushin:2011gz}. A detailed discussion of its cosmological implication, including the perturbation theory, is presented in the paper \cite{Barbashov:2005hu}. Quantum features arising within the model are discussed in the following section.

\section{Quantum Features of CCGR}\label{quantum_CCGR}

We should start with a brief discussion of various approaches to quantum gravity. The discussion is due to us attributing our results to one particular quantum gravity approach, and we briefly discuss features that may appear in others. Nowadays, the number of different approaches is huge, but one can distinguish three of the best-known directions within that landscape, namely: string theory, Loop Quantum Gravity (LQG) and the standard perturbative approach.

String theory introduces the new fundamental notion of a (super) string. Therefore, one is obliged to find the gravity theory on that notion and may adopt the results of this paper only in the low energy regime. Putting it differently, our results may serve as a way to verify the low energy string theory limit, but cannot be implemented directly in the theory's foundation.

The framework of LQG is valuable in quantum gravity study, as it is an attempt to construct a self-consistent framework of canonical quantum gravity \cite{Rovelli:1997yv,Smolin:2004sx}. The approach shares many similarities with CCGR, because LQG is founded on the Hamiltonian vierbein formalism. In the LQG framework, one introduces a dreibein (triad) connecting spatial indices with $SU(2)$ indices to cast gravity in terms of Ashtekar variables. This allows one to use the Wilson loop to solve Hamiltonian constraints (Wheeler--DeWitt equation) and build a space of state basis in the spin network formalism. In our paper, in full analogy with LQG, we adopt the vierbein (that is reduced to the dreibein at the level of ADM foliation) to connect spatial indices with a nonlinear representation of $\SL$ and $C(1,3)$ groups. In other words, we introduce new symmetry in the model, which may affect both the Wheeler--DeWitt equation and the structure of the spin network. However, this issue requires a separate treatment, which lies beyond the scope of this paper.

The standard perturbative approach was developed in the papers \cite{DeWitt:1967yk,DeWitt:1967ub,DeWitt:1967uc}, and it treats gravity as small perturbations over the Minkowski background. Although it is possible to construct a renormalizable gravity model \cite{Stelle:1976gc}, general relativity appears to be non-renormalizable in the second order of perturbation theory \cite{Goroff:1985th}. Thus, it is usually used in the low energy regime within the effective field theory framework {(see the following works for detailed reviews \cite{Buchbinder:1992rb,Burgess:2003jk})}. 

 {The direct implication of the standard perturbative approach in the context of CCGR is impossible because of the following reasons.} First, it uses the full metric $g_{\mu\nu}$ as a dynamical variable, {while CCGR admits} the vierbein as dynamical variables carrying a nonlinear symmetry representation. Thus, the perturbative approach should be stated in terms of vierbein variables. Second, in CCGR, one cannot use Minkowski spacetime as a background. {We express the cosmological epoch through the mean dilaton value, which is expressed in terms of the metric. Therefore, metric perturbations can induce cosmological expansion. The Minkowski background, on the other hand, does not admit cosmological expansion, so it cannot be used as a proper background.} In such a way, in order to implement the perturbative approach, one should define graviton states in terms of the vierbein with respect to a cosmological background.

 {In this section, we demonstrate that one can construct a consistent approach to quantum CCGR in the low energy regime. Such an approach was discussed briefly and obscured in previous papers \cite{Arbuzov:2009zza,Arbuzov:2010fz,Pervushin:2011gz}. 
One finds a solution of classical field equations describing a solitary (nonlinear) plain wave propagating in the conformal space and associates it with a single graviton state in the quasiclassical regime. Then, one decomposes the gravitational field in a series of (nonlinear) plain waves and obtains a quasiclassical graviton Lagrangian.}

 {A similar method cannot be used in general relativity because of the following. Within GR, one treats the metric as a dynamical variable, and it enters the Lagrangian in a nonlinear way. Therefore, one needs to associate quasiclassical gravitons with metric perturbations with nonlinear dynamics. In CCGR, one must use the vierbein and forms $\omega^R$ as dynamical variables carrying nonlinear symmetry representation. Forms $\omega^R$, unlike the metric, enters the gravity Lagrangian in a bilinear way due to symmetry properties, thereby allowing one to avoid nonlinear dynamics. In other words, nonlinear representation of the new symmetry allows one (at least in the quasiclassical regime) to choose basic states with bilinear dynamics.}
%

 {To implement this program, one need first to choose} a proper background for perturbation theory. As we connected time with the cosmological epoch, we can only adopt the Friedman--Robertson--Walker (FRW) spacetime. In a contrast to the standard approach, we do not consider the scale factor to be a classical (non-quantum) well-defined function, {but expect it to be defined by a dynamical quantum perturbation}. The mean part of the dilaton $\langle D \rangle$ is related to the scale factor $a$ in the following way:
\begin{eqnarray}
 \langle D \rangle = -\ln a .
\end{eqnarray}

  {We use this formula as a definition of the classical scalar factor in terms of quantum metric perturbations.} This allows us to treat gravitons at the level of the conformal metric {in a way similar to the standard perturbation approach}.

 {The next step is to find} a classical nonlinear plain wave solution {and associate it with a quasiclassical graviton state in quantum theory. First, we obtain such a solution and discuss its role in the theory later}.

The conformal metric and the dilaton are dependent variables connected by the symmetry \eqref{conformal_symmetry_on_vierbein}, so one can fix a gauge of the conformal metric by the following:
\begin{eqnarray}\label{lichnerowicz_gauge}
 \gamma=1 ~,
\end{eqnarray}
\textls[-20]{where $\gamma$ is the conformal three-metric determinant. The gauge is known as the Lichnerowicz gauge~\cite{Lichnerowicz,York:1971hw,kuchar}}. Next, we set the shift vector to be zero $N^a =0$ by a proper coordinate redefinition (see Equation \eqref{normal_laps_and_shift} from Appendix \ref{appendix_ADM_foliation}, which relates the four-coordinate frame to the shift vector). We also consider a time parameterization with $N=1$. One can use the fact that any two-dimensional space can be equipped with conformal coordinates \cite{Chandrasekhar:1985kt}, to foliate the three-space into a series of two-spaces. Putting it differently, any two-dimensional metric can be transferred to the conformal form:
\begin{eqnarray}
 g_{11} (dx^1)^2 + g_{22} (dx^2)^2 + 2 g_{11} dx^1 dx^2 = \Omega\left[ (dX^1)^2 + (dX^2)^2 \right] ~,
\end{eqnarray}
so one can cast any three-metric in the following form:
\begin{eqnarray}
 g_{33} (dx^3)^2 + 2 g_{13} dx^1 dx^3 + 2 g_{23} dx^2 dx^3 + \Omega \left[ (dx^1)^2 + (dx^2)^2 \right] ~.
\end{eqnarray}

Metric function $g_{33}$ can be made equal to one by proper $x^3$ rescaling. Metric functions $g_{13}$, $g_{23}$ are similar to the shift vector in ADM foliation, as they describe two-coordinate shift from layer to layer. Therefore, we require $g_{13}=g_{23}=0$, as in a plain-like wave, an observer must experience the gravitational gradient between two-layers (wavefront), but not within a two-layer. With the use of that factorization, we define a nonlinear plain wave solution by the following metric ansatz:
\begin{eqnarray}\label{plain_wave_metric}
 g = - d\chi^0 \otimes d\chi^0 + d\chi^3 \otimes d\chi^3 + e^\Sigma \left[ e^\sigma d\chi^1 \otimes d\chi^1 + e^{-\sigma} d\chi^2\otimes d\chi^2 \right] ~,
\end{eqnarray}
where $\sigma$ depends on $\chi^3$ and $\chi^0$, while $\Sigma$ depends only on $\chi^1$ and $\chi^2$. This metric describes a three-space foliated in a series of two-surfaces with the same topology, but their geometry varies with time. Each two-surface serves as a wavefront, while a normal vector, which is co-directed with $\chi^3$, serves as a wave-vector. In such a metric parameterization, two-surfaces can have non-flat geometry, and we use functions $\sigma$ and $\Sigma$ to distinguish two conformal factors. The factor $\Sigma$ maps the two-surface geometry, while $\sigma$ maps geometry variations with layers and time. Within the ADM foliation, the action \eqref{decoupled_Lagrangian_Graviton} reads (see Appendix \ref{appendix_ADM_foliation} for notations):
\begin{eqnarray}
 S_\text{Gravitons} =\int d\chi^0 N_0 ~\int d^3 \chi \mathcal{N}\left[ G^{abcd} K_{ab} K_{cd} + \sqrt{\gamma} \widetilde{R}^{(3)} \right]
\end{eqnarray}

Evaluated on the metric \eqref{plain_wave_metric}, it takes the following form:
\begin{eqnarray}\label{wave_action}
 S_\text{Wave}&=& \int d\chi^0 ~d^3 \chi ~\left\lbrace \cfrac12 \left[ \left( \cfrac{\pd\sigma}{\pd \chi^0} \right)^2 - \left(\cfrac{\pd\sigma}{\pd \chi^3}\right)^2 \right] \nonumber \right. \\
& & \hspace{4cm} \left. - e^{-\Sigma} \left( e^{-\sigma} \cfrac{\pd^2 \Sigma}{\pd (\chi^1)^2}+ e^\sigma\cfrac{\pd^2 \Sigma}{\pd (\chi^2)^2}\right) \right\rbrace ~.
\end{eqnarray}
 
 {The former terms in \eqref{wave_action} contain derivatives of $\sigma$; thus, they describe the influence of wavefront geometry, but do not influence wavefront dynamics. For the sake of simplicity, we consider wavefronts to have a plain topology $\Sigma=1$, as this does not affect the wave dynamics. The first two terms in \eqref{wave_action} have the form of the wave equation and describe wave dynamics given by $sigma$. Therefore, the metric \eqref{plain_wave_metric} is subjected to the wave equation and must be considered as a nonlinear plain wave.}


 {One can prove that general relativity admits a similar metric as a solution of the vacuum field equation. Thus, the metric \eqref{plain_wave_metric} describes plain waves both in general relativity and CCGR. However, one cannot associate the metric \eqref{plain_wave_metric} with the single graviton quasiclassical state within general relativity because of the following reasons. As we mentioned before, general relativity admits small metric perturbations $\delta g$ as dynamical variables and connects single graviton states to small metric perturbations. Metric \eqref{plain_wave_metric}, on the other hand, does not necessarily describe small perturbation, so one cannot associate it with a graviton state within general relativity. Moreover, one can expand the general relativity Lagrangian in series with respect to small perturbation $\delta g$, but cannot expand it with respect to the metric \eqref{plain_wave_metric}.}

 {Unlike general relativity, CCGR uses another set of dynamical variables. Thus, we need to state metric \eqref{plain_wave_metric} in terms of the vierbein and forms $\omega^R$.} ADM foliation and the vierbein formalism are related by the following formulae:
\begin{eqnarray}
 \begin{cases}\label{vierbein-ADM_relation}
 \omega_{(0)} = N dx^0 ~,\\
 \omega_{(a)}= \omega_{(b) i} \left[ dx^i + N^i dx^0 \right] ~,
 \end{cases}
\end{eqnarray}
where $\omega_{0}$, $\omega_{(i)}$ is the vierbein basis, $\omega_{(a)i}$ is the correspondent dreibein and $N$, $N_i$ is the laps function and the shift vector. One can express \eqref{vierbein-ADM_relation} in an equivalent form:
\begin{eqnarray}
 \begin{cases}\label{vierbein-ADM_relation_alternative}
 \omega_{(0)0}=N ~,\\
 \omega_{(0)i}=0 ~,\\
 \omega_{(a)0}=\omega_{(a)i} N^i ~.
 \end{cases}
\end{eqnarray}

As we showed in Section \ref{nonlinear_symmetry_representation_gravity}, the form $\omega^L$ is not a dynamical variable \eqref{metric_differential_forms}, so one needs to consider only the connection between the metric and form $\omega^R$. For the nonlinear plain wave metric \eqref{plain_wave_metric}, nonzero vierbeins are given by the following:
\begin{eqnarray}
 \omega_{(0)0}= 1,~~ \omega_{(3)3} = 1,& \omega_{(1)1} = e^{\frac12 \sigma},& \omega_{(2)2}=e^{-\frac12 \sigma},\\
 \omega_{(0)}^{~~0}=-1,~~ \omega_{(3)}^{~~3}=1,& \omega_{(1)}^{~~1}=e^{-\frac12\sigma},& \omega_{(2)}^{~~2}=e^{\frac12 \sigma} ,
\end{eqnarray}
and the element $\omega_{(\mu)\sigma} d \omega_{(\nu)}^{~~\sigma}$ is symmetric, so it matches $\omega^R$ and reads:
\begin{eqnarray}
 \omega^R_{(\mu)(\nu)}= \omega_{(\mu)\sigma} d \omega_{(\nu)}^{~~\sigma}= \cfrac12 d\sigma \left( \delta_{(\mu)(1)} \delta_{(\nu)(1)} - \delta_{(\mu)(2)} \delta_{(\nu)(2)} \right). \label{omega^R_plain_wave}
\end{eqnarray}

 {Formula \eqref{omega^R_plain_wave} gives an explicit expression for dynamical variables $\omega^R$ corresponding to a classical solitary nonlinear plain wave. Unlike metric variables, form $\omega^R$ admits the Taylor series expansion}:
\begin{eqnarray}\label{gravitons_series}
 \omega^R_{(a)(b)} (\pd_{(c)}) = \int\cfrac{d^3 k}{(2\pi)^3} \cfrac{1}{\sqrt{2 \omega_k}} ~i k_{(c)} \left[ \epsilon^R_{(a)(b)}(k) g^+_k e^{i k\cdot x}+ \epsilon^R_{(a)(b)}(-k) g^-_k e^{-i k \cdot x} \right] ~,
\end{eqnarray}
where all scalar products are evaluated on the dreibein ($x \cdot k = k_{(a)} x_{(a)}$). {Functions $\epsilon^R_{(a)(b)}(k)$ correspond to from $\omega^R$ evaluated on a solitary nonlinear plain wave with wave vector $k$; and functions $g^\pm_k$ play the role of Fourier coefficients. In such a way, Formula \eqref{gravitons_series} allows one to expand an arbitrary classical gravitational field in a series of nonlinear plain waves.} Similar to a weak wave in general relativity, $\epsilon_{(a)(b)}(k)$ is constrained by the following identities:
\begin{eqnarray}\label{polarization_constraint}
 \begin{cases}
 \epsilon^R_{(a)(a)}(k)=0 ~,\\
 k_{(a)} \epsilon^R_{(a)(b)}(k)=0 ~.
 \end{cases}
\end{eqnarray}

A single nonlinear wave is driven by the action \eqref{wave_action}, thereby fixing gravitons on a zero-mass shell, in full agreement with the classical theory:
\begin{eqnarray}
 k_\mu k^\mu =0 ~.
\end{eqnarray}

 {This allows one to associate $g^\pm_k$ with the creation and annihilation operators of quasiclassical gravitons. We would like to highlight that operators $\hat{g}^\pm_k$ should be associated with quasiclassical graviton states, i.e., with graviton states in the low energy regime. As we used a classical theory to construct the quantum model, one has no grounds to use this in the high energy regime without further analysis, which is beyond the scope of this paper.}

 {In order to describe dynamics of (conformal) gravitons in quantum CCGR, one needs to evaluate the action \eqref{decoupled_Lagrangian_Graviton} with the use of the expansion \eqref{gravitons_series}. The curvature scalar $R$ is bilinear with respect to external curvature $K_{ab}$ (see Appendix \ref{appendix_ADM_foliation}), while external curvature is linear with respect to form $\omega^R$:}
\begin{eqnarray}
 K_{ab} = \cfrac{1}{N} \left[ \omega^R_{(m)(n)} (\pd_t) ~\omega_{(m)a} \omega_{(n)b} -\cfrac12 \left(\nabla_a N_b + \nabla_b N_a\right) \right] ~.
\end{eqnarray}
 
 {Therefore, (conformal) gravitons in quantum CCGR are obliged to have bilinear dynamics. Action \eqref{decoupled_Lagrangian_Cosmology} defined the cosmological dynamics of the model, while \eqref{decoupled_Lagrangian_Potential} the driving interaction between global (cosmological) and local physics.}

One can summarize these results as follows. {One can find the classical CCGR solution describing the nonlinear plain wave. CCGR dynamical variables $\omega^R$ admit expansion in Fourier series with respect to solitary nonlinear plain waves \eqref{gravitons_series}. Thus, one can associate the plain wave solution \eqref{plain_wave_metric} with a single graviton state, and the series \eqref{gravitons_series} should be viewed as an expansion in single graviton states. Action \eqref{decoupled_Lagrangian_Graviton} defines bilinear graviton dynamics, while the actions \eqref{decoupled_Lagrangian_Cosmology} and \eqref{decoupled_Lagrangian_Potential} define cosmological dynamics and gravity self-interaction correspondingly. Therefore, we obtain a new way to perform perturbative quantization of gravity with the use of new symmetry in the low energy regime. The same method cannot be applied in general relativity, because it uses a different set of dynamical variables. Form $\omega^R$ is defined with respect to a nonlinear symmetry representation \eqref{omega^R} and cannot be adopted within general relativity.}
\vspace{6pt} 
\acknowledgments{We are grateful to Prof. V.N. Pervushin for inspiring ideas.}
\authorcontributions{The authors contributed equally to this work.}
\conflictsofinterest{The authors declare no conflict of interest.} 
\appendixtitles{yes} 
\appendixsections{multiple} 

\appendix

\section{Algebras of $A(4)$ and $C(1,3)$}\label{appendix_algebra}

The group $A(4)$ is a semidirect product of the special linear group $\SL$ and the translation group. $A(4)$ is given by shift generators $P_{(\mu)}$ that form the translation group, asymmetric (Lorentz group) generators $L_{(\mu)(\nu)}$ and symmetric generators $R_{(\mu)(\nu)}$. Symmetric and asymmetric generators form the special linear group $\SL$. Non-zero commutators are given by the following:
\begin{eqnarray}
 \cfrac{1}{i} [L_{(\mu)(\nu)},L_{(\sigma)(\rho)}] &=& \delta_{(\mu)(\sigma)} L_{(\nu)(\rho)} - \delta_{(\mu)(\rho)} L_{(\nu)(\sigma)} - ((\mu)\leftrightarrow (\nu)) ~, \\
 \cfrac{1}{i} [L_{(\mu)(\nu)},R_{(\sigma)(\rho)}] &=& \delta_{(\mu)(\sigma)} R_{(\nu)(\rho)} + \delta_{(\mu)(\rho)} R_{(\nu)(\sigma)} - ((\mu)\leftrightarrow (\nu)) ~, \\
 \cfrac{1}{i} [R_{(\mu)(\nu)},R_{(\sigma)(\rho)}] &=& \delta_{(\mu)(\sigma)} L_{(\rho)(\nu)} + \delta_{(\mu)(\rho)} L_{(\sigma)(\nu)} + ((\mu)\leftrightarrow (\nu)) ~, \\
 \cfrac{1}{i} [L_{(\mu)(\nu)},P_{(\sigma)}] &=& \delta_{(\mu)(\sigma)} P_{(\nu)} - ((\mu)\leftrightarrow (\nu)) ~,\\
 \cfrac{1}{i} [R_{(\mu)(\nu)},P_{(\sigma)}] &=& \delta_{(\mu)(\sigma)} P_{(\nu)} + ((\mu)\leftrightarrow (\nu)) ~.
\end{eqnarray}

The conformal group algebra $C(1,3)$ consists of Lorentz group generators $L_{(\mu)(\nu)}$, shift generators $P_{(\mu)}$, scaling transformation generator $D$ and special conformal transformation generators $K_{(\mu)}$. Non-zero group commutators read:
\begin{eqnarray}
[L_{(\mu)(\nu)},D]=0, & [P_{(\mu)},D] = -i P_{(\mu)}, \\
~[K_{(\mu)},K_{(\nu)}]=0, & [K_{(\mu)},D] = i K_{(\mu)}, 
\end{eqnarray}
\begin{eqnarray}
~[P_{(\mu)},K_{(\nu)}]=2 i (\delta_{(\mu)(\nu)} D - L_{(\mu)(\nu)}) , \\
~ [L_{(\mu)(\nu)},K_{(\rho)}] = i \delta_{(\mu)(\rho)} K_{(\nu)} - (\mu \leftrightarrow \nu) .
\end{eqnarray}

\section{Conformal Metric Transformations}\label{appendix_conformal_transformations}
Conformal metric transformations are defined {at the level of the metric} $g_{\mu\nu}$:
\begin{eqnarray}
 g_{\mu\nu} \to \Omega^2 g_{\mu\nu} ~,
\end{eqnarray}
where $\Omega$ is a function that performs conformal mapping. {One can use them to relate the standard (Einstein) metric $g_{\mu\nu}$ to the conformal metric} $\widetilde{g}_{\mu\nu}$:
\begin{eqnarray}
 g_{\mu\nu} = \Omega^2 \widetilde{g}_{\mu\nu} ~.
\end{eqnarray}

The following formulae \cite{Hawking:1973uf} for conformal transformations in $D$-dimensional spacetime are used:
\begin{eqnarray}
g_{\mu\nu} = \Omega^2 \widetilde{g}_{\mu\nu} ~, & g^{\mu\nu} = \Omega^{-2} \widetilde{g}^{\mu\nu} ~, & \sqrt{\lvert g\rvert} = \Omega^D \sqrt{\lvert \widetilde g \rvert} ~,
\end{eqnarray}
\begin{eqnarray}
R &=& \Omega^{-2} \widetilde R - 2(D-1) \Omega^{-3} \square \Omega - (D-1)(D-4) \Omega^{-4} (\nabla\Omega)^2 ~.
\end{eqnarray}

For a particular case of four-dimensional spacetime, the formulae are given by the following:
\begin{eqnarray}
\sqrt{\lvert g\rvert} &=& \Omega^4 \sqrt{\lvert\widetilde g\rvert} ~,\\
R &=& \Omega^{-2} \widetilde R - 6 \Omega^{-3} \square \Omega ~, \\
R\sqrt{\lvert g\rvert} &=& \sqrt{\lvert\widetilde g\rvert}\left[ \widetilde R \Omega^2-6~ \Omega\square \Omega\right] .
\end{eqnarray}

\section{Vierbein Formalism}\label{appendix_vierbein}

Vierbeins are constrained by the following orthogonality relations:
\begin{eqnarray}
 \goe_{\mu m} \goe^{\nu m} = \delta_\mu^\nu , & \goe_{\mu m} \goe^{\mu n} = \delta_m^n ~, \label{orthogonality_vierbein} \\
 \goe_{\mu m} \goe_\nu^{~~m} = g_{\mu\nu} ~ ,& \goe_{\mu m} \goe^\mu_{~~n} = \eta_{mn} ~, \nonumber
\end{eqnarray}
where $g_{\mu\nu}$ is the metric and $\eta_{mn}$ is the Minkowski metric. The following expression connects the vierbein and the metric determinant:
\begin{eqnarray}
 \sqrt{-g} = \det \goe_\mu^{~~m} ~.
\end{eqnarray}

There are many ways to define the connection in the vierbein formalism, but they are equivalent up to a complex factor. We use the following definition:
\begin{eqnarray}
 \nabla_{\mu} \phi_i = \pd_\mu \phi_i -\cfrac{i}{2} ~\left(\omega_\mu\right)^{ab} ~\left(L_{ab}\right)_i^{~j} \phi_j ~,
\end{eqnarray}
where $\phi_i$ is a field of an arbitrary spin, $\left( \omega_\mu\right)^{ab}$ is the connection and $L_{ab}$ is the Lorentz group generator in a representation suitable for the field $\phi_i$. One uses the standard definition of the Riemann tensor in the vierbein formalism:
\begin{eqnarray}
 [\nabla_\mu, \nabla_\nu] \phi_i &=& \cfrac{i}{2} (R_{\mu\nu})_i^{~j} \phi_j ~,\\
 (R_{\mu\nu})_i^{~j} &=& \pd_\mu (\omega_\nu)_i^{~j} - \pd_\nu (\omega_\mu)_i^{~j} + (\omega_\mu)_i^{~k} (\omega_\nu)_k^{~j} - (\omega_\nu)_i^{~k} (\omega_\mu)_k^{~j} ~.
\end{eqnarray}

\section{ADM Foliation}\label{appendix_ADM_foliation}

The ADM foliation formalism is {presented} in the papers \cite{Arnowitt:1962hi,Dirac:1958sc}, and a brief introduction to the formalism may be found in the book \cite{Kiefer:2004gr}.

A four-dimensional spacetime can be represented as a foliation of space-like three-surfaces. Zero coordinate $x^0$ is considered as a time coordinate, and for any given moment of time, $x^1$, $x^2$ and $x^3$ set a coordinate frame {on} a three-surface. Each surface is equipped with a metric $\gamma_{ab}$ (Latin indices take values of $1,2,3$), a normal vector $n^\mu$ and three tangent vectors $e_a$. Three-dimensional quantities' indices must be raised and lowered by the three-metric $\gamma_{ab}$. One introduces a scalar function $N$ called the laps and a vector $N_a$ called the shift vector. They are defined by the following expression:
\begin{eqnarray}
 \cfrac{\pd x^\mu}{\pd x^0}= N n^\mu + N^a e^\mu_a ~, \label{normal_laps_and_shift}
\end{eqnarray}
where $e^\mu_a$ are {four-coordinates} of a tangent vector $e_a$. The four-dimensional metric is given by the~following:
\begin{eqnarray}
g_{\mu\nu} &=& \begin{pmatrix} N_a N^a - N^2 & N_b \\ N_c & \gamma_{ab} \end{pmatrix} ~,\\
g^{\mu\nu} &=& \begin{pmatrix} -\cfrac{1}{N^2} & \cfrac{N^b}{N^2} \\ & \\ \cfrac{N^b}{N^2} & \gamma^{ab} - \cfrac{N^a N^b}{N^2} \end{pmatrix} ~,\\
\sqrt{-g} &=& N \sqrt{\gamma} ~.
\end{eqnarray}

Three-surfaces are immersed in the four-dimensional spacetime {and have two geometrical characteristics}. The (internal) curvature is defined by the standard Riemann {tensor}:
\begin{eqnarray}
{}^{(3)}R_{mn~b}^{~~~a}=\pd_{m} \Gamma^a_{nb} - \pd_n \Gamma^a_{mb} + \Gamma^a_{ms} \Gamma^s_{nb} - \Gamma^a_{ns} \Gamma^s_{mb} ~.
\end{eqnarray}

The external curvature $K_{ab}$ is given by the following expression:
\begin{eqnarray}
K_{ab} = \cfrac{1}{2N} \left( \dot{\gamma}_{ab} -\nabla_a N_b -\nabla_b N_a \right) ~.
\end{eqnarray}

The following equation connects the four-dimensional curvature $R$ with the three-dimensional curvature ${}^{(3)} R$:
\begin{eqnarray}
R={}^{(3)} R+\cfrac{1}{\sqrt{\gamma}} G^{abcd} ~K_{ab} K_{cd} ~,
\end{eqnarray}
where $G^{abcd}$ is the DeWitt supermetric:
\begin{eqnarray}
G^{abcd} &=&\cfrac{\sqrt{\gamma}}{2} \left( \gamma^{ac} \gamma^{bd} + \gamma^{ad} \gamma^{bc} - 2 \gamma^{ab} \gamma^{cd} \right) ~, \\
G_{abcd} &=& \cfrac{1}{2\sqrt{\gamma}} \left( \gamma_{ac}\gamma_{bd}+\gamma_{ad}\gamma_{bc}-\gamma_{ab}\gamma_{cd} \right) ~, \\
G^{abcd}G_{cdef} &=& \cfrac12 \left( \delta^a_e \delta^b_f + \delta^b_e \delta^a_f \right) ~.
\end{eqnarray}

\reftitle{References}


\end{document}